\documentclass[10pt,a4paper]{article}
\usepackage[latin1]{inputenc}
\usepackage{amsmath}
\usepackage{amsfonts}
\usepackage{amssymb}
\usepackage{makeidx}
\usepackage{graphicx}
\usepackage{subfig}
\usepackage[left=2.00cm, right=2.00cm, top=2.00cm, bottom=2.00cm]{geometry}
\author{Soham Chandra$^{1,a)}$ and Muktish Acharyya$^{1,b)}$ \medskip \\ \textit{${}^{1}$Department of Physics, Presidency University, Kolkata- 700 073}\vspace{20pt}\\ ${}^{a)}$Corresponding author: soham.rs@presiuniv.ac.in\\${}^{b)}$muktish.physics@presiuniv.ac.in}
\title{\large\textbf{A Monte Carlo Study on the Variation of Residual Magnetisation with the Ratio of Coupling Strengths and Non-magnetic Impurities in an Ising Trilayer}}
\begin{document}
	\date{}
	\maketitle
	\vspace{30pt}
	\begin{abstract}
		We have studied a spin-1/2, ABA, Ising trilayer system with two different types of interactions, in-plane ferromagnetic and out-of-plane anti-ferromagnetic, among the lattice sites, on different layers. In the pure case, devoid of any impurity, we employed Monte-Carlo method with single spin-flip Metropolis algorithm to find out the anti-ferromagnetic critical temperature (N\'{e}el temperature) and another lower temperature, called the compensation temperature, both with total
		magnetisation zero, in accordance with already established results for square lattice. Then non-magnetic impurities, (spin value$=0$) were
		implanted on each layer at randomly picked sites and their concentration was increased in steps from 5$\%$ to 20$\%$. The ratio
		of inter-planar anti-ferromagnetic to mid-layer ferromagnetic coupling strength as well as the ratio of top and bottom layer
		ferromagnetic to mid layer ferromagnetic coupling strength were also varied and N\'{e}el and compensation temperatures,
		both were observed shifting towards lower temperature values with increase in concentration of impurities in the lattice for
		any fixed ratio of different coupling strengths, in absence of any external magnetic field. In addition, the magnitude of the
		residual magnetisation i.e. the ratio of the peak value of the magnetisation in between N\'{e}el and compensation points and
		the saturated value of magnetisation, was also observed to vary with different values of controlling parameters.
	\end{abstract}
	\vspace{40pt}
	\section*{\centering{INTRODUCTION}}
	\indent Magneto-caloric effect (MCE) in magnetic systems, is rigorously investigated due to its potential for applications in cryogenics and construction of energy efficient devices [1-4]. Recent analytical advancements to the theory of MCE mostly use scaling-based equations of state, using mean field approximation for constructing the thermodynamics of
	magnetic systems [5-11]. On the front of numerical analysis, Monte Carlo (MC) simulations [12] are used to predict the magneto-caloric properties of materials [13-15]. There are quite a few prominent exactly solvable spin models, exploiting various approaches for which magneto-caloric quantities have been discussed, such as the Jordan-Wigner transformation [16], Bethe ansatz-based quantum transfer matrix and nonlinear integral equations method [17].\\
	\indent Magnetic multilayers have attracted significant attention recently from technologists and theoreticians alike for	the fact that the properties of the multilayer system as a whole can significantly be different from those of any of its component layers. This is the motivation behind the study of them from both the schools. Thus we've witnessed many
	interesting observations and unusual phase diagrams from both the communities for a magnet when it is composed of	materials with different interactions, e.g. ferromagnetic and anti-ferromagnetic. Due to advancement of experimental techniques the properties of the multilayer may be tailored as per need by preparing the multilayer structures	synthetically. To meet desired macroscopic characteristics we can vary the microscopic structure, e.g. by implanting
	impurity in the layering pattern.\\
	\indent Ferrimagnetic layered materials are exciting for the existence of compensation points, i.e., temperatures below the anti-ferromagnetic critical point or N\'{e}el temperature for which the total magnetization is zero while each of the individual layers still remain magnetically ordered [18]. Although unrelated to critical phenomena (till date, not proven
	otherwise), some physical properties of the system (e.g., the magnetic coercivity) may exhibit a singular behavior at the compensation point [19-21]. In [21] the authors have found out the compensation point of some ferrimagnets occurs near room temperature which makes them candidates for magneto-optical drives. A quasi three-dimensional, spin-1/2, ABA, Ising trilayer stacking with quenched non-magnetic impurity (spin value $=0$) is studied in [22] and it is observed that both, the N\'{e}el and compensation points, can be modulated to drift towards lower temperatures with increasing concentration of non-magnetic impurities.\\
	\indent This article uses an already established model and its thermodynamic description [18, 22]. The controlling
	parameters for the presence of magnetic compensation phenomenon are varied accordingly to find out the gradual	change in the ratio of the peak value of magnetisation between N\'{e}el and compensation temperatures and the saturated value of magnetisation at the lowest value of temperature in the simulation scheme which we wish to call residual magnetisation. The numerical results are extensively illustrated in tables and plots.
	\section*{\centering{MODEL AND INTERACTION HAMILTONIAN}}
	We performed our numerical investigation of the magnetic and thermodynamic properties on a spin-1/2 Ising trilayer system containing three magnetic layers with the following constructional details:
	\begin{itemize}
		\item[(a)] Each layer is composed exclusively either of the two possible types of atoms, A or B.
		\item[(b)] There are three types of interactions or bonds:\\
		A-A $\to$ Ferromagnetic\\
		B-B $\to$ Ferromagnetic\\
		A-B $\to$ Anti-ferromagnetic.
		\item [(c)] The layers can be stacked, distinguishably, in two ways as (i) ABA stacking and (ii) AAB stacking (Fig.1).
	\end{itemize}
	\begin{figure}[ht]
		\centering
		\subfloat[AAB stacking] {\includegraphics[width=0.45\textwidth]{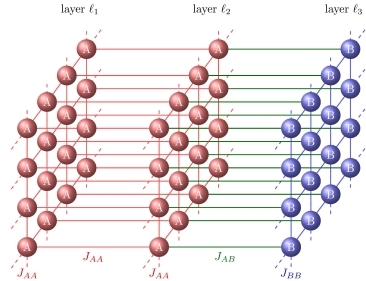}\label{fig:f1}}
		\hfill
		\subfloat[ABA stacking] {\includegraphics[width=0.45\textwidth]{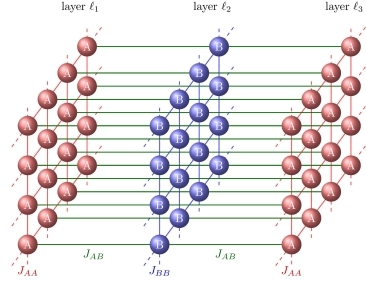}\label{fig:f2}}
		\caption{Two distinct trilayer stackings: (a) AAB stacking (b) ABA stacking. Image courtesy [18]}
	\end{figure}
	The equilibrium spin configuration on any layer is determined by the condition that the free energy of the whole system should be a minimum. We consider here two contributions to the total free energy coming from, the in-plane spin-spin interactions of the three layers and out-of-plane spin-spin interactions of top-mid and mid-bottom layers. As we've considered the spins to interact Ising-like, in-plane as well as out-of-plane, we can write the Hamiltonian for the ABA trilayer system as:
	\begin{equation}
	H=-J_{11}\sum_{t}S_{t}S_{t+1}-J_{22}\sum_{m}S_{m}S_{m+1}-J_{33}\sum_{b}S_{b}S_{b+1}-J_{12}\sum_{<t,m>}S_{t}S_{m}-J_{23}\sum_{<m,b>}S_{m}S_{b}
	\end{equation}
	where summation indices $t, m$ and $b$ denote respectively the lattice site indices for the top-layer, mid layer and bottom-layer. In equation $(1)$, the first, second and third terms respectively are for the intra-planar contributions from
	the top-layer, mid-layer and bottom-layer. The fourth and the fifth terms are the contributions arising out of the nearest neighbor inter-planar interactions, between top and mid layers and mid and bottom layers.
	The nature of the coupling strengths are: $J_{11}<0$ , $J_{22}<0$, $J_{33}<0$ and $J_{12}>0$, $J_{23}>0$ supporting ferromagnetic intraplanar and anti-ferromagnetic inter-planar interactions. We've considered periodic boundary conditions in-plane
	(i.e. along both the mutually perpendicular directions residing on any of the planes) and open boundary conditions along the vertical, so that there is no out-of-plane interaction term between the top and bottom layer in the Hamiltonian.
	
	\section*{\centering{SIMULATION SCHEME AND CALCULATED QUANTITIES}}
	We performed Monte Carlo simulations [20] with single spin-flip Metropolis algorithm of the system above using the following steps. At a fixed temperature $T$, we choose a lattice site $i$ randomly and flip the spin value, from $S_{i}$ to $-S_{i}$ by using the Metropolis rate [10]:
	\begin{equation}
	P(S_{i} \to -S_{i}) = \text{min} (1, \exp (\Delta E/k_{B}T))
	\end{equation}	
	where $\Delta E$ is the change of energy due to the change of the orientation of the spin projection from $S_{i}$ to $-S_{i}$. We set	the Boltzmann constant, $k_{B}$ to $1$. Similar $3L^{2}$ random updates of spins are defined as one Monte Carlo sweep per spin
	(MCSS). Corresponding to a high-temperature, we started from an initially random configuration of the trilayer and equilibrated the system till $12\times10^{4}$ MCSS and calculated thermal averages and fluctuations from further $4\times10^{4}$ MCSS.
	So, the total MCSS in the simulation for a fixed temperature, $T$ is $16\times10^{4}$ . Moving forward we decrease the temperature	and use the last spin configuration at the previous higher temperature as the initial configuration for the new lower temperature. This procedure simulates a cooling that is closer to equilibrium compared to starting at each temperature	with a random spin configuration. The CPU time needed for $16\times10^{4}$ MCSS is approximately $18$ h on an Intel$^\circledR$ Core$^{\text{TM}}$ 
	i5-6500 CPU @ 3.20 GHz$\times$4, for a trilayer with $128\times128$ sites on each layer.
	We are interested in tracking down the changes in \textit{residual magnetisation}, of the system as a whole for different	parameter values with changes in temperature from $3.1$ to $0.1$. We have chosen to vary the following parameters in the ways described below for each fixed value of temperature:
	\begin{itemize}
		\item[1.] We have chosen three values of $J_{11}/J_{22}$ as $0.20, 0.35, 0.50$ and set $J_{33}=J_{11}$ , $J_{23}=J_{12}$ .
		\item[2.] For each value of $J_{11}/J_{22}$ , we varied the ratio of $J_{12}/J_{22}$ to $-0.1, -0.4, -0.7$.
		\item[3.] For each of the values of $J_{11}/J_{22}$ and $J_{12}/J_{22}$ , we've calculated the ensemble averages of the following quantities at each of the temperature points:\\
		\textbf{(1) Sublattice magnetisations} for top, mid and bottom layers, denoted by $M_{t}$ , $M_{m}$ , $M_{b}$ by the following formulae:
		\begin{equation}
		M_{q}=\frac{1}{3L^{2}}\sum_{x,y=1}^{L} \left<S_{x,y}\right>
		\end{equation}
		where $q \in \{t,m,b\}$ and the sum extends over all sites in respective planes as x and y denote the co-ordinates of	a spin on a plane and runs from $1$ to $L=128$, after completion of one TMCS. $\langle\cdots\rangle$ denotes an average over time i.e. after completion of MCSSs, after equilibration. We assume ergodicity has been reached so an ensemble	average equals the time average.\\
		\textbf{(2) Total magnetisation of the trilayer:} $M=M_{t}+M_{m}+M_{b}$
		\item[4.] From the plot of $M$ vs. temperature, we find the values of residual magnetisation and study the effects of controlling parameters on it.
		\item[5.] Then we repeated the above procedure for randomly implanted non-magnetic impurity concentrations (spin value $=0$) of 5\%, 10\%, 15\% and 20\% and observed the variation of residual magnetisation on impurity concentration.
	\end{itemize}
		
	\section*{\centering{NUMERICAL RESULTS AND DISCUSSION}}
	\textbf{TABLE 1.} Variation of residual magnetization, obtained for the pure trilayer case, with the variation of controlling parameters namely $J_{11} /J_{22}$ , $J_{12}/J_{22}$ and concentration of implanted non-magnetic (spin value $=0$) impurities, $C$ (n.d. implies not detected):\medskip\\
	\begin{center}
	\begin{tabular}{|c|c|c|c|c|c|c|}
		\hline 
		\rule[-1ex]{0pt}{4.5ex} $J_{11}/J_{22}$& $J_{12}/J_{22}$ & C=0.00 (pure) & C=0.05 & C=0.10 & C=0.15 & C=0.20  \\ 
		\hline 
		
		\rule[-1ex]{0pt}{3.5ex}  	 & -0.10 & -0.782 & -0.770 & -0.763 & -0.746 & -0.721 \\ 
		 
		\rule[-1ex]{0pt}{3.5ex} 0.20 & -0.40 & -0.696 & -0.676 & -0.663 & -0.634 & -0.594 \\ 
			
		\rule[-1ex]{0pt}{3.5ex}  	 & -0.70 & -0.565 & -0.536 & -0.511 & -0.467 & -0.409 \\ 
		\hline 
		\rule[-1ex]{0pt}{3.5ex}      & -0.10 & -0.394 & -0.373 & -0.355 & -0.326 & -0.290 \\ 
		 
		\rule[-1ex]{0pt}{3.5ex} 0.35 & -0.40 & -0.250 & -0.222 & -0.202 & -0.170 & -0.125 \\ 
		 
		\rule[-1ex]{0pt}{3.5ex}      & -0.70 & -0.087 & -0.063 & -0.045 & -0.018 & -0.001 \\ 
		\hline 
		\rule[-1ex]{0pt}{3.5ex}      & -0.10 & -0.143 & -0.121 & -0.106 & -0.080 & -0.038 \\ 
		 
		\rule[-1ex]{0pt}{3.5ex} 0.50 & -0.40 & -0.023 & n.d. & n.d. & n.d. & n.d. \\ 
		
		\rule[-1ex]{0pt}{3.5ex}      & -0.70 & n.d. & n.d. & n.d. &  n.d. & n.d. \\ 
		\hline 
	\end{tabular} 
	\end{center}
	\newpage 
	\begin{figure}[h]
		\centering
		\includegraphics[width=1.0\linewidth]{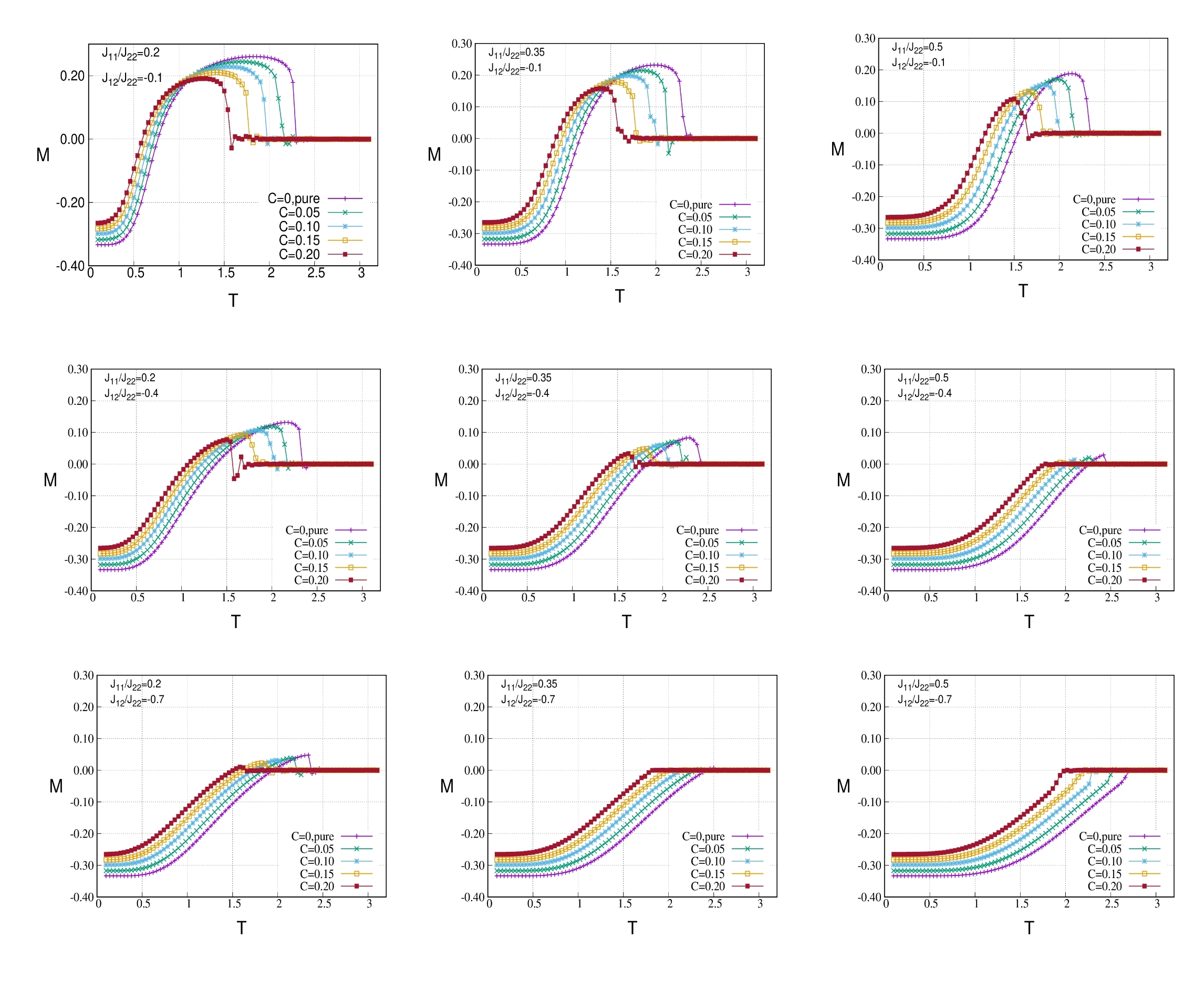}
		\caption{Plots of variation of residual magnetization with the variation of controlling parameters namely $J_{11}/J_{22}, J_{12}/J_{22}$ and	concentration of implanted non-magnetic impurities}
		\label{fig:allcombined}
	\end{figure}

	\section*{\centering{CONCLUSION}}
	\indent We studied the spin-1/2, ABA, Ising trilayer, with MC simulations with single spin-flip Metropolis algorithm, first without any impurity to find out the residual magnetisation decreases as the ratio of intra-planar ferromagnetic coupling strengths of top layer to mid-layer increase with a fixed value of inter-planar antiferromagnetic to mid-layer
	ferromagnetic ratio. Similar trend in residual magnetisation is observed while we kept decreasing the value of inter-planar antiferromagnetic to mid-layer ferromagnetic ratio, after fixing the ratio of intra-planar ferromagnetic coupling	strengths of top layer to mid-layer.\\
	\indent Next we moved on with the same scheme but implanted random, non-magnetic impurities (spin value $=0$) to observe decay in the value of residual magnetisation with increasing impurity concentration from 5\% to 20\%, for both the above scenarios i.e. two types of variation of ratios of coupling strengths. We found out that for fixed ratios, increase in impurity concentration results in gradual decrease in the value of residual magnetisation.	
	%\vspace{30pt}
	\section*{\centering{ACKNOWLEDGEMENTS}}
	We would like to gratefully acknowledge the Computational facilities and Library resources provided by	Presidency University. S.C. and M.A. would like to extend their gratitude respectively towards University Grants Commission (UGC) and FRPDF grant from Presidency University for financially sponsoring their efforts. S.C. would like to also thank departmental colleagues Sangita Bera, Tamaghna Maitra and Sukhendu Mukherjee for helping in formatting the entire article as per need.
	
	\section*{\centering{REFERENCES}}
	\begin{itemize}
		\item [1.] Spichkin Y I and Tishin A M, The Magnetocaloric Effect and Its Applications (Institute of Physics Publishing,
		Philadelphia, 2003).
		\item [2.]Tishin A and Spichkin Y, Int. J. Refrigeration 37, 223 (2014).
		\item [3.]Gschneidner K A Jr, Pecharsky V K and Tsokol A O, Rep. Prog. Phys. 68, 1479 (2005).
		\item [4.]Szymczak H and Szymczak R, Mater. Sci. Poland 26, 807 (2008).
		\item [5.]Pe\l{}ka R et al., Acta Phys. Pol. A 124, 977 (2013).
		\item [6.]Franco V, Conde A, Romero-Enrique J M and Bl\'{a}zquez J S, J. Phys.: Condens. Matter 20, 285207 (2008)
		\item [7.]Amaral J S and Amaral V S, Chapter 8 in Thermodynamics: Systems in Equilibrium and Non-Equilibrium, edited
		by J C Moreno-Piraj\'{a}n, pp. 173?198 (2011).
		\item [8.]Amaral J S, Silva N J O and Amaral V S, Appl. Phys. Lett. 91, 172503 (2007).
		\item [9.]de Oliveira N and von Ranke P, Phys. Rep. 489, 89 (2010).
		\item [10.]Dong Q Y, Zhang H W, Sun J R, Shen B G and Franco V, J. Appl. Phys. 103, 116101 (2008).
		\item [11.]Basso V, Sasso C P and K\"{u}pferling M, Int. J. Refrigeration 37, 257 (2014).
		\item [12.]K. Binder, D.W. Heermann, Monte Carlo simulation in statistical physics (Springer, New York, 1997).
		\item [13.]N\'{o}brega E P, de Oliveira N A, von Ranke P J and Troper A, Phys. Rev. B 72, 134426 (2005).
		\item [14.]N\'{o}brega E P, de Oliveira N A, von Ranke P J and Troper A, J. Magn. Magn. Mater. 310, 2805 (2007).
		\item [15.]Singh N and Arr\'{o}yave R, J. Appl. Phys. 113, 183904 (2013).
		\item [16.]Topilko M, Krokhmalskii T, Derzhko O and Ohanyan V, Eur. Phys. J. B 85, 1 (2012).
		\item [17.]Trippe C, Honecker A, Kl\"{u}mper A and Ohanyan V, Phys. Rev. B 81, 054402 (2010).
		\item [18.]I. J. L. Diaz and N. S. Branco, Physica B 73, 529 (2017).
		\item [19.]G. Connell, R. Allen, and M. Mansuripur, Journal of Applied Physics 53, 7759 (1982).
		\item [20.]H. P. D. Shieh and M. H. Kryder, Applied physics letters 49, 473 (1986).
		\item [21.]J. Ostorero, M. Escorne, A. Pecheron-Guegan, F. Soulette, and H. Le Gall, Journal of Applied Physics 75, 6103
		(1994).
		\item [22.]Sk. Sajid and Muktish Acharyya, Phase Transition (2019) in press. arXiv:1907.07879v1.
	\end{itemize}	 
\end{document}